\documentclass{elsart}

 \usepackage{graphicx}

\usepackage{amssymb}
\usepackage{bm}

\newcommand{\dto}{Dy$_{2}$Ti$_{2}$O$_{7}$}
\newcommand{\hto}{Ho$_{2}$Ti$_{2}$O$_{7}$}

\begin{document}

\begin{frontmatter}

\title{Spin correlation in kagom\'{e} ice state: neutron scattering study of 
the dipolar spin ice \dto\ under magnetic field along [111]}

\author[aff1]{Y. Tabata}
\author[aff2]{H.~Kadowaki}
\author[aff3]{K.~Matsuhira}
\author[aff4]{Z.~Hiroi}
\author[aff5]{N.~Aso}
\author[aff6]{E.~Ressouche}
\author[aff6]{B.~F\aa k}
\address[aff1]{Graduate School of Science, Osaka University, Toyonaka, Osaka 560-0043, Japan}
\address[aff2]{Department of Physics, Tokyo Metropolitan University, Hachioji-shi, Tokyo 192-0397, Japan}
\address[aff3]{Department of Electronics, Faculty of Engineering, Kyushu Institute of Technology, Kitakyushu 804-8550, Japan}
\address[aff4]{Institute for Solid State Physics, University of Tokyo, Kashiwa, Chiba 277-8581, Japan}
\address[aff5]{NSL, Institute for Solid State Physics, University of Tokyo, Tokai, Ibaraki 319-1106, Japan}
\address[aff6]{CEA, D\'{e}partement de Recherche Fondamentale sur la Mati\'{e}re Condens\'{e}e, SPSMS, 38054 Grenoble, France}


\begin{abstract}
We have investigated the kagom\'{e} ice state 
in the frustrated pyrochlore oxide {\dto} 
under magnetic field along a [111] axis. 
Spin correlations have been measured by neutron scattering and 
analyzed by Monte-Carlo simulation. 
The kagom\'{e} ice state, which has a non-vanishing residual entropy 
well established for the nearest-neighbor spin ice model by the exact 
solution, has been 
proved to be realized in the dipolar-interacting spin ice {\dto} by 
observing the characteristic spin correlations. 
The simulation shows that the long-range 
interaction gives rise to only weak lifting of the ground 
state degeneracy and that the system freezes within the 
degenerate kagom\'{e} ice manifold .
\end{abstract}

\begin{keyword}
geometrical frustration \sep spin ice \sep kagom\'{e} ice \sep spin correlations \sep neutron scattering
\PACS 75.50.-y \sep 75.40.Cx \sep 75.25.+z \sep 75.40.Mg
\end{keyword}

\end{frontmatter}

The spin ice model, an Ising ferromagnet with a 
local $<$111$>$ easy axis on the pyrochlore lattice 
interacting with nearest-neighbor exchange coupling, has 
attracted much attention because 
of its intriguing frustration mechanism 
\cite{Bramwell01}.
It has macroscopically degenerate ground states 
characterized by the ``two-in and two-out'' local structure 
on each tetrahedron, 
equivalent to the ``ice rule'' in the water ice 
\cite{Anderson_theory}. 
The spin ice behavior, e.g. the residual entropy \cite{Ramirez_DTO}, 
was discovered in the rare earth pyrochlore oxides, 
{\hto} \cite{Harris_HTO}, {\dto} \cite{Ramirez_DTO} and 
Ho$_2$Sn$_2$O$_7$ \cite{kado}. 
However in these systems Ising spins on the rare earth sites interact 
via a long-range dipolar interaction, and the mechanism showing 
the spin ice behavior in these dipolar spin ice was puzzling. 
A numerical account was provided by 
the Monte-Carlo(MC) simulation studies \cite{Bramwell01}, 
showing that the ground state manifold is approximately preserved 
in the dipolar model. 
More recently an elegant explanation using an analytical 
method was proposed in a theoretical work \cite{Isakov_mode}. 

The macroscopic degeneracy of the spin ice model is partly or 
fully lifted by magnetic fields \cite{spin-ice_field}. 
Along a [111] axis the pyrochlore lattice can be viewed 
as an alternating stacking of kagom\'{e} and triangular 
layers. For field along the [111] axis, the degeneracy is 
partly lifted, where random spin configurations are retained 
only in the kagom\'{e} planes, and has been studied by 
an exactly solvable dimer model \cite{Moessner_kagome,Udagawa_kagome}. 
Surprisingly, this kagom\'{e} ice state based on the nearest-neighbor 
interacting model, was suggested to be observed experimentally 
in the dipolar spin ice compound 
\cite{Matsuhira_kagome,Hiroi_resS}. 

Unlike the case in zero field, the effects of the long-range dipolar interaction have not been clarified in magnetic field, and hence it is still puzzling whether the degeneracy of the kagom\'{e} ice manifold is preserved in the dipolar spin ice. To resolve this question, we have measured static spin correlations of the dipolar spin ice compound \dto\ in magnetic field along the [1 1 1] axis by neutron scattering and observed characteristic spin correlations. By comparing with MC-simulation results based on the dipolar model, we have demonstrated that the kagom\'{e} ice state really occurs in the dipolar spin ice.

\begin{figure}[t]
 \begin{center}
 \includegraphics[scale =0.6]{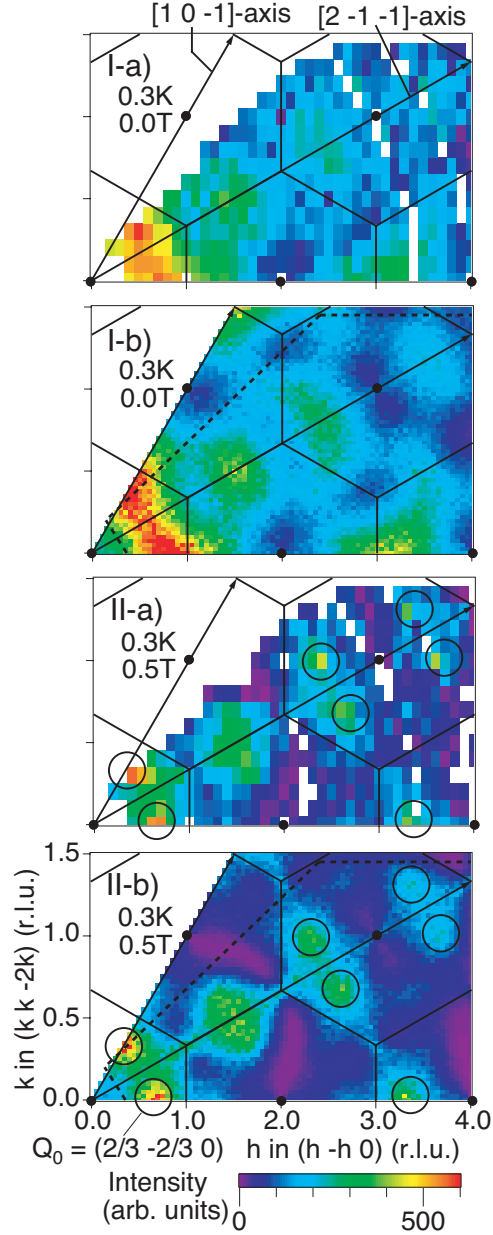}
 \end{center}
 \caption{Neutron intensity maps in the perpendicular plane to the [1 1 1] axis. Experimental results (I-a) at $T$ $=$ 0.3 K and $H$ $=$ 0.0 T and (II-a) at $T$ $=$ 0.3 K and $H$ $=$ 0.5 T. MC-simulation results (I-b) at $T$ $=$ 0.3 K and $H$ $=$ 0.0 T and (II-b) at $T$ $=$ 0.3 K and $H$ $=$ 0.5 T. Energy parameters used for the simulations in the dipolar model are $J_{1}$ $=$ -3.72 K, $J_{2}$ $=$ 0.1 K, $J_{3}$ $=$ -0.03 K and $D$ $=$ 1.41 K, where the 1st, 2nd, 3rd nearest neighbor exchange interactions and the dipolar coupling constant respectively. The values of $J_{1}$, $J_{3}$ and $D$ are assigned to those reported previously\cite{Hertog_PRL,Ruff_inH} and the value of $J_{2}$ is determined to reproduce the temperature dependences of the specific heat in finite fields\cite{Tabata_cond-mat}.}
 \label{map}
\end{figure}

Neutron scattering experiments were performed using the triple-axis spectrometer GPTAS installed at JRR-3M JAERI. Two single crystalline samples of \dto , which were grown by the floating-zone method using an infrared furnace, were used for the experiments. The samples were mounted in a dilution refrigerator. MC-simulations were carried out using the facilities of the Supercomputer Center at ISSP University of Tokyo, which utilize the standard Metropolis single spin flip algorithm.

We show neutron intensity maps in the scattering plane perpendicular to the $[1 \, 1 \, 1]$ axis in zero and finite($H$ $=$ 0.5 T) fields at $T$ $=$ 0.3 K obtained from the experiments and MC-simulations based on the dipolar model in Fig.~\ref{map}. Experimental results show definite differences between the spin correlations at $H$ $=$ 0.0 and 0.5 T, as shown in Fig.~\ref{map} (I-a) and (II-a) respectively. The characteristic correlations at ${\bm Q}_{0}$ $=$ $($$2/3$, $-2/3$,  $0$$)$(${\bm Q}_{0}$-correlation), the positions are represented by black circles in (II-a), is found at $H$ $=$ 0.5 T.  

The MC-simulations based on the dipolar model well reproduce the neutron intensity map at $H$ $=$ 0.5 T(Fig.~\ref{map} (II-b)), as well as that at $H$ $=$ 0.0 T(Fig.~\ref{map} (I-b)). The characteristic ${\bm Q}_{0}$-correlation observed in the experiment is clearly reproduced in Fig.~\ref{map} (II-b). The ${\bm Q}_{0}$-correlation was proposed by the theoretical work based on the nearest-neighbor model\cite{Moessner_kagome}, and hence, it is a good signature of the kagom\'{e} ice state. The agreement between the experimental and the MC-simulation results on the static spin correlations, especially the observation of the ${\bm Q}_{0}$-correlations, indicates that the degeneracy of the kagom\'{e} ice manifold is approximately preserved in the dipolar spin ice \dto . Spin correlations at $H$ $=$ 0.5 T do not show significant temperature dependence below 0.75 K in both the experiments and the single spin flip MC-simulation. It strongly suggests that the system freezes within the degenerate kagom\'{e} ice manifold.

In summary, we have presented the investigation of the neutron scattering experiments of the dipolar spin ice compound \dto\ in the magnetic field along the $[1 \, 1 \, 1]$ axis. The analysis of the experimental results by the single spin flip MC-simulations reveals that the kagom\'{e} ice state is realized in the dipolar spin ice.

\end{document}